# Quantum and Classical Variance in the Quantum Realm

**Mario Rabinowitz**[1]

___

This paper examines the variance of quantum and classical predictions in the quantum realm, as well as unexpected presence and absence of variances. Some features are found that share an indirect commonality with the Aharonov-Bohm and Aharonov-Casher effects in that there is a quantum action in the absence of a force. Variances are also found in the presence of a force that are more subtle as they are of higher order. Significant variances related to the harmonic oscillator and particle in a box periods are found. This paper raises the question whether apparent quantum self-inconsistency may be examined internally, or must be empirically ascertained. These inherent variances may either point to inconsistencies in quantum mechanics that should be fixed, or that nature is manifestly more non-classical than expected. For the harmonic oscillator it is proven that $\langle x^2 \rangle_{QM} = \langle x^2 \rangle_{CM}$.

___

## 1. INTRODUCTION

Quantum and classical variances will be examined, as well as the unexpected presence and absence of variances. The variances related to the harmonic oscillator and particle in a box periods are noteworthy as they persist as the quantum number $n \to \infty$. These variances seem not to have been previously analyzed, and appear to be both prevalent and experimentally testable. Similarities are found with the Aharonov-Bohm and Aharonov-Casher effects in that there is a quantum action in the absence of a force.

Therefore these effects will be discussed quantum mechanically and classically to facilitate comparison with the effects found in this paper.

It is tempting to start with a particle in a box with perfectly reflecting walls as this is a fundamental problem with the simplest solutions for the wave functions. It also has an advantage in comparing with a classical particle since the wave function is completely contained inside the box. However, some may raise questions about the infinite gradient of the potential energy (infinite force) at the walls and non-locality as the source of the variances.

[1]Armor Research; 715 Lakemead Way; Redwood City, CA 94062-3922 USA, e-mail Mario715@gmail.com

On the other hand when variances are found for a particle with a finite force acting on it such as the simple harmonic oscillator (SHO), they may be ascribed to penetration of the wave function into the classically forbidden regions. The reason for the incompatibility is different in the two cases.

## 2. AHARONOV-BOHM, AHARONOV-CASHER EFFECTS, AND BERRY'S PHASE

The Aharonov-Bohm[1] and Aharonov-Casher[2] effects are commonly thought to be explainable only by quantum mechanics (QM). Even Berry's geometric phase seems amenable to classical interpretation. It is not the purpose of this section to side with either the quintessential quantum, or classical explanations, but this will be by way of contrast, as the quantum-classical expectation value variances presented in this paper are not the result of electric or magnetic fields, or due to phase differences; and appear not to have classical explanations.

### 2.1 Aharonov-Bohm Effect

The question of which is more fundamental, force or energy is central to the foundations of physics, though it is somewhat rendered void in the Lagrangian or Hamiltonian formulations. In Newtonian classical mechanics (CM), force (*vis motrix* in Newton's Principia[3]), and kinetic energy (*vis viva* in Leibnitz' Acta erud.[4], ) are two of the foremost concepts. In QM, potential and kinetic energies are the primary concepts, with force hardly playing a role at all. It was not until 1959, some thirty-three years after the advent of QM that Aharonov and Bohm described gedanken electrostatic and magnetostatic cases in which physically measurable effects occur where presumably no forces act.[1] This is now known as the Aharonov-Bohm (A-B) effect.

In the magnetic case, an electron beam is sent around both sides of a long shielded solenoid or toroid so that the electron paths encounter no magnetic field and hence no magnetic force. Electrons do encounter a magnetic vector potential, which enters into the electron canonical momentum producing a phase shift of the electron wave function, and hence QM interference. If the electrons go through a double slit and screen apparatus the shielded magnetic field shifts the interference pattern periodically as a function of $h/e$ in the shielded region, where $h$ is Planck's constant and $e$ is the electronic charge (in superconductors because of electron pairing, the magnetic flux quantum is $h/2e$).

This was confirmed experimentally and considered a triumph for QM. The A-B effect appears not to have been seriously challenged for forty-one years until 2000 when Boyer [5,6] argued that the A-B effect can be understood completely classically. First he points out that there has been no real experimental confirmation of the A-B effect. The periodic phase shift of a two-slit interference pattern due to a shielded magnetic field has indeed been confirmed. However, no experiment has shown that there are no forces on the electrons, that the electrons do not accelerate, and that the electrons on the two sides of a solenoid (or toroid) are not relatively displaced.

Boyer then goes on to propose a classical mechanism. The electron induces a field in the conductor (shield or electromagnet) and this field acts back on the charged particle producing a force which speeds up the particle as it approaches and then slows the particle as it recedes, so that it time averages to 0. This sequence is reversed on the other side of the magnetic source giving interference. The displaced charge in the shield (or solenoid windings) affects the current in the solenoid, and hence the center-of-energy of the solenoid field.

**2.2 Aharonov-Casher effect**

In 1984 Aharonov-Casher[2] (A-C) proposed an analog of the A-B effect in which the electrons are replaced by neutral magnetic dipoles such as neutrons, and the shielded magnetic flux is replaced by a line charge. They claimed that the neutral magnetic dipole particles undergo a quantum phase shift and show an effect despite experiencing no classical force. The A-C effect has been confirmed experimentally, and although it is considered to be solely in the domain of QM, Boyer also proposed a classical interpretation of this effect.

In 1987 Boyer[7] argued that neutrons passing a line charge experience a classical electromagnetic force in the usual electric-current model for a magnetic dipole. This force will produce a relative lag between dipoles passing on opposite sides of the line charge, with the classical lag leading to a quantum phase shift as calculated by A-C. Boyer went on to predict that a consequence of his analysis is the breakdown of the interference pattern when the lag becomes comparable to the wave-packet coherence length.

In 1991, Mignani[8] showed that the A-C effect is a special case of geometrical phases, i.e. the standard Berry phase and the gauge-invariant Yang phase.

### 2.3 Berry's Geometric Phase

In 1984, the same year as the A-C effect, Berry[9] theoretically discovered that when an evolving quantum system returns to its original state, it has a memory of its motion in the geometric phase of its wavefunction. There are both quantum and classical examples of Berry's geometric phase (BGP), but as far as I know no one has yet challenged the QM case with a CM explanation. It is noteworthy that in 1992 Aharonov and Stern[10] did the QM analog of Boyer's[7] CM analysis, in examining BGP in terms of Lorentz-type and electric-type forces to show that BGP is analogous to the A-B effect.

### 3. SIMPLE HARMONIC OSCILLATOR (SHO)

To avoid the possibility that the classical and quantum variances shown here are in any way related to any kind of electromagnetic forces, we shall deal only with neutral particles that have no electric or magnetic moments. But nevertheless, they have a variance in their classical and quantum position expectation values. Of course it could be argued that most, if not all neutral particles are composed of charged constituents.

## 3.1 Classical Harmonic Oscillator

We begin with the classical harmonic oscillator so that we may compare with the corresponding expectation values for a quantum harmonic oscillator. Let us normalize the classical probability density P which is inversely proportional to the oscillating particle's velocity

$$1 = \int_{-A}^{A} \frac{b}{\pm w(A^2 - x^2)^{1/2}} dx \Rightarrow b = \frac{\pm w}{p}, \tag{3.1}$$

where b is the normalization constant, A is the classical amplitude, and the angular frequency $w = 2pu$. Therefore the normalized classical probability density is

$$bP = \frac{1}{p(A^2 - x^2)^{1/2}}. \tag{3.2}$$

The classical particle position expectation values are

$$\langle x \rangle_{CM} = \int_{-A}^{A} x \left[ \frac{1}{p(A^2 - x^2)^{1/2}} \right] dx = 0, \tag{3.3}$$

and all $\langle x^k \rangle_{CM} = 0$ for odd values of k = 1, 3, 5, … because P is even and $x^k$ is odd for all odd k.

$$\langle x^2 \rangle_{CM} = \int_{-A}^{A} x^2 \left[ \frac{1}{p(A^2 - x^2)^{1/2}} \right] dx = \frac{A^2}{2}. \tag{3.4}$$

$$\langle x^4 \rangle_{CM} = \int_{-A}^{A} x^4 \left[ \frac{1}{p(A^2 - x^2)^{1/2}} \right] dx = \frac{3A^4}{8}. \tag{3.5}$$

$$\langle x^6 \rangle_{CM} = \int_{-A}^{A} x^6 \left[ \frac{1}{p(A^2 - x^2)^{1/2}} \right] dx = \frac{5A^6}{16}. \tag{3.6}$$

## 3.2 Quantum Harmonic Oscillator

The time independent Schrödinger equation for the SHO for a particle of mass m, oscillating with frequency $f$, and angular frequency $w = 2pf$, is:

$$\frac{-(h/2p)^2}{2m}\nabla^2 y + (2p^2 mf^2 x^2)y = Ey \tag{3.7}$$

The eigenfunction solution to Eq. (3.7) for the one-dimensional SHO is

$$y_n(x) = b_n e^{-\frac{x^2}{2}} H_n(x) = b_n e^{-\frac{a^2 x^2}{2}} H_n(ax), \tag{3.8}$$

where n = 0, 1, 2, 3,…,  $x \equiv ax, a \equiv 2p[Mf/h]^{1/2} = [2pMw/h]^{1/2}$, and $H_n(x)$ is the Hermite polynomial of the $n$th degree in $x$:

$$H_n(x) = (-1)^n e^{x^2} \frac{d^n e^{-x^2}}{dx^n}. \tag{3.9}$$

In general, the normalization constant

$$b_n = \left[\frac{a}{p^{1/2} 2^n n!}\right]^{1/2}. \tag{3.10}$$

We shall use the quantum energy level solution to the classical energy

$$E_n = (n + \tfrac{1}{2})hf = (n + \tfrac{1}{2})h(w/2p) = (\tfrac{1}{2})mw^2 A^2 \tag{3.11}$$

to help in the comparison of the classical and quantum position expectation values.

### *3.2.1 Ground State n = 0 for Harmonic Oscillator*

Let us examine the ground state expectation values $<x^k>_{QM}$ where the variance with classical mechanics (CM) is expected to be the greatest here. The normalized eigenfunction for the ground state (n = 0) is

$$y_0(x) = \frac{a^{1/2}}{p^{1/4}} e^{-\frac{a^2 x^2}{2}}. \tag{3.12}$$

In general, the expectation value of $<x^k>_{QM0}$ is

$$\left\langle x^{k}\right\rangle_{QM\,0}=\int_{-\infty}^{\infty}y_{0}^{*}x^{k}y_{0}dx=\int_{-\infty}^{\infty}x^{k}\left[\frac{a^{1/2}}{p^{1/4}}e^{-\frac{a^{2}x^{2}}{2}}\right]^{2}dx. \tag{3.13}$$

The expectation value of $<x^k>_{QM} = 0$ for odd values of the index k = 1, 3, 5, …. because $y_0(x)$ is an even function and $x^k$ is odd. In general $<x^k>_{QM} = <x^k>_{CM} = 0$, and in particular $<x>_{QM} = <x>_{CM} = 0$ by symmetry in QM and CM.

$$\left\langle x\right\rangle_{QM\,0}=\int_{-\infty}^{\infty}x\left[\frac{a^{1/2}}{p^{1/4}}e^{-\frac{a^{2}x^{2}}{2}}\right]^{2}dx=0=\left\langle x\right\rangle_{CM}. \tag{3.14}$$

So let us focus on some even values of k.

$$\left\langle x^{2}\right\rangle_{QM\,0}=\int_{-\infty}^{\infty}x^{2}\left[\frac{a^{1/2}}{p^{1/4}}e^{-\frac{a^{2}x^{2}}{2}}\right]^{2}dx=\frac{1}{2a^{2}}=\frac{A^{2}}{2}=\left\langle x^{2}\right\rangle_{CM}. \tag{3.15}$$

$$\left\langle x^{4}\right\rangle_{QM\,0}=\int_{-\infty}^{\infty}x^{4}\left[\frac{a^{1/2}}{p^{1/4}}e^{-\frac{a^{2}x^{2}}{2}}\right]^{2}dx=\frac{3}{4a^{4}}=\frac{3A^{4}}{4}=2\left\langle x^{4}\right\rangle_{CM}. \tag{3.16}$$

$$\left\langle x^{6}\right\rangle_{QM\,0}=\int_{-\infty}^{\infty}x^{6}\left[\frac{a^{1/2}}{p^{1/4}}e^{-\frac{a^{2}x^{2}}{2}}\right]^{2}dx=\frac{15}{8a^{6}}=\frac{15A^{6}}{8}=6\left\langle x^{6}\right\rangle_{CM}. \tag{3.17}$$

### 3.2.2 First Excited State n = 1 for Harmonic Oscillator

$$\left\langle x\right\rangle_{QM\,1}=\int_{-\infty}^{\infty}x\left[\frac{a^{1/2}}{2^{1/2}p^{1/4}}(2ax)e^{-\frac{a^{2}x^{2}}{2}}\right]^{2}dx=0=\left\langle x\right\rangle_{CM}. \tag{3.18}$$

$$\left\langle x^{2}\right\rangle_{QM\,1}=\int_{-\infty}^{\infty}x^{2}\left[\frac{a^{1/2}}{2^{1/2}p^{1/4}}(2ax)e^{-\frac{a^{2}x^{2}}{2}}\right]^{2}dx=\frac{3}{2a^{2}}=\left\langle x^{2}\right\rangle_{CM}. \tag{3.19}$$

$$\left\langle x^{4}\right\rangle_{QM\,1}=\int_{-\infty}^{\infty}x^{4}\left[\frac{a^{1/2}}{2^{1/2}p^{1/4}}(2ax)e^{-\frac{a^{2}x^{2}}{2}}\right]^{2}dx=\frac{15}{4a^{4}}=\frac{10}{9}\left\langle x^{4}\right\rangle_{CM}. \tag{3.20}$$

$$\left\langle x^{6}\right\rangle_{QM\,1}=\int_{-\infty}^{\infty}x^{6}\left[\frac{a^{1/2}}{2^{1/2}p^{1/4}}(2ax)e^{-\frac{a^{2}x^{2}}{2}}\right]^{2}dx=\frac{105}{8a^{6}}=\frac{14}{9}\left\langle x^{6}\right\rangle_{CM}. \tag{3.21}$$

### 3.2.3 Second Excited State n = 2 for Harmonic Oscillator

$$\left\langle x\right\rangle_{QM\,2}=\int_{-\infty}^{\infty}x\left[\frac{a^{1/2}}{2p^{1/4}2^{1/2}}(4a^{2}x^{2}-2)e^{-\frac{a^{2}x^{2}}{2}}\right]^{2}dx=0=\left\langle x\right\rangle_{CM}. \tag{3.22}$$

$$\left\langle x^2 \right\rangle_{QM2} = \int_{-\infty}^{\infty} x^2 \left[ \frac{a^{1/2}}{2p^{1/4}2^{1/2}} (4a^2x^2 - 2)e^{-\frac{a^2x^2}{2}} \right]^2 dx = \frac{5}{2a^2} = \left\langle x^2 \right\rangle_{CM}. \tag{3.23}$$

$$\left\langle x^4 \right\rangle_{QM2} = \int_{-\infty}^{\infty} x^4 \left[ \frac{a^{1/2}}{2p^{1/4}2^{1/2}} (4a^2x^2 - 2)e^{-\frac{a^2x^2}{2}} \right]^2 dx = \frac{39}{4a^4} = \frac{26}{25} \left\langle x^4 \right\rangle_{CM}. \tag{3.24}$$

$$\left\langle x^6 \right\rangle_{QM2} = \int_{-\infty}^{\infty} x^6 \left[ \frac{a^{1/2}}{2p^{1/4}2^{1/2}} (4a^2x^2 - 2)e^{-\frac{a^2x^2}{2}} \right]^2 dx = \frac{375}{8a^6} = \frac{6}{5} \left\langle x^6 \right\rangle_{CM}. \tag{3.25}$$

### 3.3 Comparison of Quantum and Classical Harmonic Oscillator

We now compare the quantum and classical harmonic oscillator position expectation values based upon Eqs. (2.4) to (2.6), and (2.14) to (2.25). As proven below, it is noteworthy that $\left\langle x^2 \right\rangle_{CM} = \left\langle x^2 \right\rangle_{QM}$, although all higher order position even moments are not equal; and of course $\left\langle x^k \right\rangle_{QM} = \left\langle x^k \right\rangle_{CM} = 0$ for all odd k = 1, 3, 5, …. The higher order CM position even moments are significantly smaller than the higher order QM position even moments, and the disparity increases as the moments get larger. This can be attributed to penetration of the quantum wave function into the classically forbidden region for both even and odd $y_n(x)$ as $y^*y = |y^2|$ is even and enters into the integration. This effect will diminish as one goes to higher quantum states, and should disappear as $n \to \infty$ for pure states. It is not clear that this will happen for wave packets.[22]

As shown earlier in Eq. (3.8) $y_n(x) = b_n e^{-\frac{x^2}{2}} H_n(\mathbf{x}) = b_n e^{-\frac{a^2x^2}{2}} H_n(ax)$, where $\mathbf{x} \equiv ax$, $\mathbf{a} \equiv 2p[Mf/h]^{1/2} = [2pMw/h]^{1/2}$.

$$\left\langle \mathbf{x}^2 \right\rangle_{QM} = \int_{-\infty}^{\infty} y_n^* \mathbf{x}^2 y_n dx = \int_{-\infty}^{\infty} y_n^* \left[ \tfrac{1}{2}(\mathbf{x} + \tfrac{d}{d\mathbf{x}}) + \tfrac{1}{2}(\mathbf{x} - \tfrac{d}{d\mathbf{x}}) \right]^2 y_n dx$$

$$= \int_{-\infty}^{\infty} y_n^* \left[ \tfrac{1}{4}(\mathbf{x} + \tfrac{d}{d\mathbf{x}})^2 + \tfrac{1}{4}(\mathbf{x} - \tfrac{d}{d\mathbf{x}})^2 + \tfrac{1}{4}(\mathbf{x} + \tfrac{d}{d\mathbf{x}})(\mathbf{x} - \tfrac{d}{d\mathbf{x}}) + \tfrac{1}{4}(\mathbf{x} - \tfrac{d}{d\mathbf{x}})(\mathbf{x} + \tfrac{d}{d\mathbf{x}}) \right] y_n dx \tag{3.26}$$

$$= \int_{-\infty}^{\infty} y_n^* \left[ \tfrac{1}{4}(\mathbf{x} + \tfrac{d}{d\mathbf{x}})^2 + \tfrac{1}{4}(\mathbf{x} - \tfrac{d}{d\mathbf{x}})^2 + \tfrac{1}{2}(-\tfrac{d^2}{d\mathbf{x}^2} + \mathbf{x}^2) \right] y_n dx$$

For the SHO:

$$\left\langle PotentialEnergy \right\rangle_{QM} = \left\langle PE \right\rangle_{QM} = \tfrac{1}{2} M w^2 \left\langle x^2 \right\rangle, \tag{3.27}$$

$$\left( \mathbf{x} + \tfrac{d}{d\mathbf{x}} \right)^2 y_n = 2[n(n-1)]^{1/2} y_{n-2}, \tag{3.28}$$

$$\left(x - \tfrac{d}{dx}\right)^2 y_n = 2[(n+1)(n+2)]^{1/2} y_{n+2}, \text{ and} \tag{3.29}$$

$$\int_{-\infty}^{\infty} y_n y_j \, dx = 0 \text{ for } n \neq j \tag{3.30}$$

because the Hermite polynomials are orthogonal, leaving only the 3rd term of the integrand in Eq. (3.26). Substituting, $x \equiv \alpha x$ and multiplying Eq. (3.26) by $(h\omega/4\pi)$:

$$\begin{aligned}
\langle PE \rangle_{QM} &= \tfrac{1}{2} M \omega^2 \langle x^2 \rangle_{QM} = \left(\frac{h\omega}{4\pi}\right) \int_{-\infty}^{\infty} y_n^* \left[\tfrac{1}{2}\left(-\left(\frac{h}{2\pi m \omega}\right)\frac{d^2}{dx^2} + \left(\frac{2\pi m \omega}{h}\right)x^2\right)\right] y_n \, dx \\
&= \int_{-\infty}^{\infty} y_n^* \left[-\left(\frac{h^2}{2(4\pi^2 M)}\right)\frac{d^2}{dx^2} + \tfrac{1}{2} M \omega^2 x^2\right] y_n \, dx \qquad\qquad (3.31) \\
&= \tfrac{1}{2} \int_{-\infty}^{\infty} y_n^* E_n y_n = \tfrac{1}{2} E_n = \tfrac{1}{2}\left(n + \tfrac{1}{2}\right)\frac{h}{2\pi}\omega
\end{aligned}$$

Since $\langle PE \rangle_{QM} + \langle KE \rangle_{QM} = E_n$, Eq. (3.31) implies $\langle PE \rangle_{QM} = \langle KE \rangle_{QM}$. Classically

$$\tfrac{1}{2} M \omega^2 \langle x^2 \rangle_{CM} = \langle PE \rangle_{CM} = \langle KE \rangle_{CM} = \tfrac{1}{2} E = \tfrac{1}{2} E_n. \tag{3.32}$$

Therefore $\langle x^2 \rangle_{QM} = \langle x^2 \rangle_{CM}$. This could also have been obtained directly from the Virial Theorem which holds both in QM and CM.

The significance of the difference in the classical and quantum higher order position moments is that Newton's Second Law of Motion is violated because the wave function penetrates the classically forbidden regions so that the particle spends less time in the central region and more time in the region of the classical turning points than allowed by Newton's Second Law. Next let us look at the opposite case where a particle spends more time in the central region because the wave function terminates at the boundary rather than penetrating it.

## 4. FREE PARTICLE IN A BOX

The infinite square well is an archetype problem of QM. It is used as a model for a number of significant physical systems such as free electrons in a metal, long molecule, the Wigner box, etc.

### 4.1 Quantum Case for Particle in a Box

The Schrödinger non-relativistic wave equation is:

$$\frac{-(h/2\pi)^2}{2m}\nabla^2 y + Vy = i(h/2\pi)\frac{\partial}{\partial t}y, \tag{4.1}$$

where $y$ is the wave function of a particle of mass m, with potential energy V. In the case of constant V, we can set V = 0 as only differences in V are physically significant. A solution of Eq. (3.1) for the one-dimensional motion of a free particle of nth state kinetic energy $E_n$ is:

$$y = b_n e^{i2\pi x/\lambda} e^{-i2\pi E_n t/h} = b_n e^{i2\pi\left(\frac{x}{\lambda} - \frac{\omega}{2\pi}t\right)}, \tag{4.2}$$

where the wave function $y$ travels along the positive x axis with wavelength $\lambda$, angular frequency $\omega$, and phase velocity $v = \lambda\omega/2\pi$.

We shall be interested in the time independent solutions. The following forms are equivalent:

$$\begin{aligned} y_n &= b_n e^{i2\pi x/\lambda} = b_n \cos(2\pi x/\lambda) + i\sin(2\pi x/\lambda) \\ &= b_n \sin(n\pi x/2a - n\pi/2) \end{aligned}, \quad n = 1, 2, 3, \ldots. \tag{4.3}$$

where we consider the particle to be in an infinite square well potential with perfectly reflecting walls at x = -a, and x = +a, so that $\frac{n}{2}\lambda = 2a$. The wall length 2a can be arbitrarily large, but needs to be finite so that the normalization coefficient is non-zero.

We normalize the wave functions to yield a total probability of finding the particle in the region -a to +a, and find

$$1 = \int_{-a}^{a} y^* y \, dx = \int_{-a}^{a} |y|^2 \, dx \Rightarrow b_n = \frac{1}{\sqrt{a}} \tag{4.4}$$

where the normalization is independent of n.

In general

$$\langle x^k \rangle = \int_{-a}^{a} y^* x^k y \, dx = \int_{-a}^{a} x^k |y|^2 \, dx, \text{ for k = 1, 2, 3, ….} \tag{4.5}$$

Since $|y|^2$ is symmetric **here** for both $y_{ns}$ and $y_{nas}$, $x^k|y|^2$ is antisymmetric in the interval -a to +a, because $x^k$ is antisymmetric. Thus without having to do the integration we know that $\langle x^k \rangle = 0$ for all odd k, and in particular $\langle x \rangle = 0$ for the nth state. Let us find the expectation values $\langle x^k \rangle$ where for k = 1, 2, 4, and 6 for the free particle in the nth state.

$$\langle x \rangle_{QM} = \int_{-a}^{a} y^* x y\, dx = \int_{-a}^{a} x|y|^2 dx = 0. \tag{4.6}$$

$$\langle x^2 \rangle_{QM} = \int_{-a}^{a} y^* x^2 y\, dx = \int_{-a}^{a} x^2|y|^2 dx = a^2\left[\frac{1}{3} - \frac{2}{p^2 n^2}\right] = \frac{a^2}{3}\left[1 - \frac{6}{p^2 n^2}\right]. \tag{4.7}$$

$$\langle x^4 \rangle_{QM} = \int_{-a}^{a} y^* x^4 y\, dx = \frac{a^4}{5} - \frac{4a^2(p^2 n^2 a^2 - 6a^2)}{p^4 n^4} = \frac{a^4}{5}\left[1 - \frac{20}{p^2 n^2} + \frac{120}{p^4 n^4}\right]. \tag{4.8}$$

$$\langle x^6 \rangle_{QM} = \int_{-a}^{a} y^* x^6 y\, dx = \frac{a^6}{7} - \frac{6a^2(120 a^4 - 20 p^2 n^2 a^4 + p^4 n^4 a^4)}{p^6 n^6} = \frac{a^6}{7}\left[1 - \frac{5040}{p^6 n^6} - \frac{720}{p^4 n^4} + \frac{42}{p^2 n^2}\right]. \tag{4.9}$$

Let us compare these values with the corresponding classical values.

### 4.2 Classical Case for Particle in a Box

The classical probability P is inversely proportional to the velocity whose magnitude is constant throughout the box (except at the walls). Therefore P is uniform for finding a classical free particle in the region -a to +a. Normalizing the classical probability,

$$1 = \int_{-a}^{a} bP\, dx = bP(2a) \Rightarrow bP = \frac{1}{2a}. \tag{4.10}$$

As for the quantum case, classically <$x^k$> = 0 for all odd k because P is an even function. The classical expectation value of <x> and <$x^2$> are

$$\langle x \rangle_{ClassicalMechanics} = \langle x \rangle_{CM} = \int_{-a}^{a} bP\, dx = \int_{-a}^{a} \frac{x}{2a} dx = 0. \tag{4.11}$$

$$\langle x^2 \rangle_{CM} = \int_{-a}^{a} x^2 bP\, dx = \int_{-a}^{a} \frac{x^2}{2a} dx = \frac{a^2}{3}. \tag{4.12}$$

$$\langle x^4 \rangle_{CM} = \int_{-a}^{a} x^4 bP\, dx = \int_{-a}^{a} \frac{x^4}{2a} dx = \frac{a^4}{5}. \tag{4.13}$$

$$\langle x^6 \rangle_{CM} = \int_{-a}^{a} x^6 bP\, dx = \int_{-a}^{a} \frac{x^6}{2a} dx = \frac{a^6}{7}. \tag{4.14}$$

### 4.3 Comparison of Quantum and Classical Cases

$$\langle x \rangle_{QM} = 0 = \langle x \rangle_{CM}. \tag{4.15}$$

$$\langle x^2 \rangle_{QM} = \left[1 - \frac{6}{\mathbf{p}^2 n^2}\right] \langle x^2 \rangle_{CM}. \tag{4.16}$$

$$\langle x^4 \rangle_{QM} = \left[1 - \frac{20}{\mathbf{p}^2 n^2} + \frac{120}{\mathbf{p}^4 n^4}\right] \langle x^4 \rangle_{CM}. \tag{4.17}$$

$$\langle x^6 \rangle_{QM} = \left[1 - \frac{5040}{\mathbf{p}^6 n^6} - \frac{720}{\mathbf{p}^4 n^4} + \frac{42}{\mathbf{p}^2 n^2}\right] \langle x^6 \rangle_{CM}. \tag{4.18}$$

It is clear from the analysis that the expectation values of all the odd moments $\langle x^k \rangle$ (k = 1, 3, 5, …) are exactly equal to 0 for both QM and CM. As one might expect, for even moments the variance between QM and CM is largest for small n, and furthermore is larger the higher the moment. It is also clear from Eqs. (4.16) to (4.18) that the QM even position moments approach the CM values as n gets large.

The result $\langle x \rangle_{QM} = 0 = \langle x \rangle_{CM}$ means that in moving with a constant velocity between the walls of a box, a particle spends an equal amount of time on either side of the box and hence the expectation value for finding it, is at the center of the box. However, the results disagree for higher order moments such as $\langle x^2 \rangle_{QM} = \left[1 - \frac{6}{\mathbf{p}^2 n^2}\right] \langle x^2 \rangle_{CM}$ for a particle in a perfectly reflecting box of length 2a between walls. At low quantum number n, this is smaller than the classical value $\langle x^2 \rangle_{CM} = \frac{a^2}{3}$ of Eq. (3.13). This implies that not only does the particle spend an equal time on either side of the origin, but that the particle spends more time near the center of the box independent of the length a. Since we can make the length *a* arbitrarily large, this effect is due to quantum mechanical non-locality of the presence of the walls making itself felt near the center of the box because it does not go away with large *a*. It is noteworthy that non-locality appears in such a fundamental case.

This is a violation of Newton's First Law of Motion (NFLM) because the particle must slow down in the region of the origin even though there is a force on it only at the walls. The particle cannot both be going at a constant velocity between the walls, slow down near the center, and speed up again as it goes toward the opposite wall even if the walls are arbitrarily long. Therefore in this example, we have a quantum action on a particle even where there is no force. This is a simpler case than the Aharonov-Bohm[1], Aharonov-Casher[2] (1984), and similar effects, has many of the same elements, and may be even more intrinsic to QM. It is noteworthy that unlike such effects, it is independent of Planck's constant h; and significantly there are no fields.

## 5 Quantum And Classical Periods

The object of this section is to relate QM phase and beat periods to CM periods.

### 5.1 Simple Harmonic Oscillator (QM Phase Period)

In general a wave packet representing a particle is given by a linear sum of the eigenfunctions for a given Hamiltonian

$$\Psi(x,t) = \sum_{n=1}^{\infty} b_n y_n(x) e^{-i\omega t} = \sum_{n=1}^{\infty} b_n y_n(x) e^{-i2\pi E_n t/h}, \tag{5.1}$$

because of the linearity of the Schrödinger equation. In particular for the simple harmonic oscillator, the energy eigenfunctions $y_n$ are given by Eq. (3.8) in terms of the Hermite polynomials. As we shall make a general argument here, it is not necessary to specify the particular eigenfunctions. We can see from Eq. (5.1) that the wave packet will complete $N$ full quantum mechanical phase periods, $Nt_{QM}$, when all the phase factors $e^{-i2\pi E_n t/h}$ are equal. Since $e^{-i2\pi E_n t/h} = \cos[2\pi E_n t/h] - i\sin[2\pi E_n t/h]$, this occurs when

$$2\pi E_n t/h = \frac{2\pi E_n N t_{QM}}{h} = 2\pi N + q, \tag{5.2}$$

where $q$ is the phase, and $N$ is an integer that may vary as a function of n. To satisfy Eq. (5.2), $q$ is either a constant, or only exceptional values of n may be used for the eigenfunctions that make up the wave packet. In the more general case $q$ = constant, so we may set $q = 0$ for convenience. Then, Eq. (5.2) implies

$$Nt_{QM} = \frac{h}{E_n}[N] \Rightarrow t_{QM} = \frac{h}{E_n}, \tag{5.3}$$

where we are effectively considering one period with $N = 1$.

Thus from Eq. (5.3), quantum mechanically the phase period for the one-dimensional SHO wave packet is

$$t_{QM} = \frac{h}{E_n} = \frac{h}{(n+\tfrac{1}{2})(h/2\pi)\omega} = \frac{2\pi}{(n+\tfrac{1}{2})\omega}. \tag{5.4}$$

Classically the period is

$$t_{CM} = \frac{1}{f} = \frac{2p}{w}.$$ (5.5)

Taking the ratio of Eqs. (5.4) and (5.5):

$$\frac{t_{QM}}{t_{CM}} = \frac{2p}{(n+\frac{1}{2})w}\left[\frac{w}{2p}\right] = \frac{1}{(n+\frac{1}{2})} \xrightarrow[n\to\infty]{} 0.$$ (5.6)

For $n = 1$, $\frac{t_{QM}}{t_{CM}} = \frac{2}{3}$, and since the ratio decreases monotonically as $n$ increases, the two phase periods are never equal, and $t_{QM} < t_{CM}$.

## 5.2 Free Particle in a Box (QM Phase Period)

The QM energy levels peculiarly get further from the CM energy levels, for a free particle in a box. The QM energy dependence is

$$E = \frac{1}{2m}[p]^2 = \frac{1}{2m}\left[\frac{h}{l}\right]^2 = \frac{1}{2m}\left[\frac{h}{4a/n}\right]^2 = \frac{h^2}{2m}\left[\frac{n^2}{16a^2}\right] = E_1 n^2.$$ (5.7)

Because these energy levels go as $n^2$ they get further apart $\left[(n+1)^2 - n^2 = 2n+1\right]$ as $n$ increases unlike the classical continuum, and also unlike position expectation levels. This is also unlike the QM harmonic oscillator and most other potentials. This violates the Correspondence Principle unless $h \to 0$ as $n \to \infty$, since the energy levels are proportional to $h^2 n^2$. Otherwise energy states get further apart, while the position variance gets closer.

This peculiarity warrants a comparison of the classical and quantum periods. Classically the period for the one-dimensional motion of a particle of velocity $v$ in a box of wall separation 2a is

$$t_{CM} = \frac{4a}{v} = \frac{4a}{\left[\frac{2E}{m}\right]^{1/2}} = 4a\left[\frac{m}{2E}\right]^{1/2}.$$ (5.8)

Now let us examine the quantum mechanical phase period. From the general argument by which Eq.(5.3) was derived for a wave packet:

$$t_{QM} = \frac{h}{E} = \frac{h}{E_n} = \frac{h}{E_1 n^2}.$$ (5.9)

Thus from Eqs. (5.2) and (5.3)

$$\frac{t_{QM}}{t_{CM}} = \frac{h}{E} \bigg/ 4a\left[\frac{m}{2E}\right]^{1/2} = \frac{h}{E}\frac{1}{4a}\left[\frac{2E}{m}\right]^{1/2} = \frac{h}{2a\sqrt{2mE_1 n^2}} = \frac{h}{2an\sqrt{2mE_1}} \xrightarrow[n\to\infty]{} 0. \quad (5.10)$$

Note that $t_{QM} > t_{CM}$ for $n = 1$; $t_{QM} = t_{CM}$ for $n = 2$, and thereafter $t_{QM} < t_{CM}$. Except for the first 2 energy states, this trend is the same as the SHO for the phase $t_{QM}$.

## 5.3 Quantum Beat Periods [Beat Period = (Beat Frequency)$^{-1}$]

It is possible that the observable periods and hence the only periods relevant for the Correspondence Principle are associated with beats between the phases for adjoining energy states, i.e. $t_{QMb} = h/(E_{n+1} - E_n)$ in general, rather than the phase period $t_{QM} = h/E_n$ [cf. eqs. (5.4) and (5.9)] which may or may not be measurable. [This is analogous to the classical difference between phase velocity (which can be superluminal) and subluminal group velocity, where $v_p v_g = c^2$. The quantum beat frequency $w_{QMb}/2\pi = (E_{n+1} - E_n)/h = h(w/2\pi)[(n+1+1/2) - (n+1/2)]/h = (w/2\pi)$ is traditionally observed, e.g. atomic spectra.] For the Simple Harmonic Oscillator:

$$\left[\frac{t_{QMb}}{t_{CM}}\right]_{SHO} = \frac{2\pi/w_{QMb}}{2\pi/w} = \frac{2\pi/w}{2\pi/w} = 1 \text{ for all n. } \textbf{(Accord with CM)} \quad (5.11)$$

In this case for the Infinite Square Well:

$$\left[\frac{t_{QMb}}{t_{CM}}\right]_{ISW} = \frac{2n}{2n+1} \xrightarrow[n\to\infty]{} 1. \quad (5.12)$$

For $n = 1$, $t_{CM} = 1.5 t_{QMb}$, and yet for the SHO $[t_{CM} = t_{QMb}]_{SHO}$ exactly for all n.

## 6. DISCUSSION

Although Quantum Mechanics (QM) is considered to be a theory that applies throughout the micro- and macro-cosmos, it has fared badly in the quantum gravity realm as discussed by Rabinowitz,[11, 12] and there is no extant theory after almost a century of effort.[13, 14] In the case of the macroscopic classical realm, it is generally believed that quantum expectation values should correspond to classical results in the limit of large quantum number n, or equivalently in the limit of Planck's constant $h \to 0$. Some processes thought to be purely and uniquely in the quantum realm like tunneling, can with proper modeling also exist in the classical realm as shown by Cohn and Rabinowitz.[15]

Bohm has long contended that classical mechanics is not a special case of quantum mechanics.[16, 17] The present paper makes an even stronger statement that the predictions of both Newton's First and Second Laws are violated in the quantum realm. So quantum mechanics is incompatible with them in that domain despite the fact that Newton's Second Law can be derived by QM.[19] Bohr's[19] Correspondence Principle formulated in 1928 argues that QM yields CM as the quantum number $n \to \infty$, though the results here for harmonic oscillator and particle in a box periods appear not to do so. This needs to be examined more closely in terms of Ehrenfest's theorem for expectation values.

Stochastic Electrodynamics (SED) was proposed by Boyer[20] as one possible alternative to QM. Boyer's theory of random electrodynamics is a classical electron theory involving Newton's equations for particle motion due to the Lorentz force, and Maxwell's equations for the electromagnetic fields with point particles as sources. Boyer introduced a background of random, classical fluctuating zero-point fields whose origin was in the initial stochastic processes of the big bang and are regenerated to the present. Boyer's theory of random electrodynamics is a classical theory that provides a link between classical theory with h = 0 and quantum electrodynamics. Boyer's more recent work suggests the possibility of an equilibrium between the zero-point radiation spectrum and matter which is universal (independent of the particle mass).[21]

## 7. CONCLUSION

The free particle in a box and the simple harmonic oscillator (SHO) are examined in detail to uncover classical and quantum variances. The results indicate that such variances may be expected to be found commonly for a wide range of quantum phenomena. Quantum mechanics gives the illusion of obeying Newton's laws in the quantum realm because it starts with a Hamiltonian that incorporates Newton's laws, and because QM can derive Newton's law (since it was formulated to do so). As shown in this paper, QM is incompatible with Newton's 1st and 2nd laws in the quantum domain, and perhaps even in the classical limit. Significant differences were found in this analysis for QM and CM expectation values. Since expectation values are supposed to correspond to possible classical measurements, one may be optimistic that these findings are amenable to experimental test. In addition to variances related to the expectation values of position moments, the results here for periods of the harmonic oscillator and particle in a box are noteworthy. Although the latter quantum results are obtained for wave packets as $n \to \infty$, this needs to be examined more closely in terms of Ehrenfest's theorem for expectation values as $n \to \infty$.

This paper raises the question whether apparent quantum self-inconsistency may be examined internally, or must be empirically ascertained. If there is an inherent lack of internal verifiability, this may either point to inconsistencies in quantum mechanics that should be fixed, or that nature is manifestly more non-classical than one would judge from the Hamiltonian used to obtain quantum solutions. The answer is not obvious.

19